\documentclass[conference]{IEEEtran}
\IEEEoverridecommandlockouts

\usepackage{cite}
\usepackage{amsmath,amssymb,amsfonts}
\usepackage{algorithmic}
\usepackage{graphicx}
\usepackage{textcomp}
\usepackage{xcolor}
\usepackage{multirow}
\def\BibTeX{{\rm B\kern-.05em{\sc i\kern-.025em b}\kern-.08em
    T\kern-.1667em\lower.7ex\hbox{E}\kern-.125emX}}

\begin{document}

\title{Analyzing Images of Blood Cells with Quantum Machine Learning Methods: Equilibrium Propagation and Variational Quantum Circuits to Detect Acute Myeloid Leukemia}


\author{\IEEEauthorblockN{Azra Bano}
\IEEEauthorblockA
{\textit{Electrical and Computer Engineering}\\
\textit{Rutgers University} \\
Piscataway, NJ, USA \\
ab2895@scarletmail.rutgers.edu}

\and
\IEEEauthorblockN{Larry S. Liebovitch*}
\IEEEauthorblockA{\textit{AC4 in the Climate School} \\
\textit{Columbia University} \\
New York, NY, USA \\
lsl2140@columbia.edu}

\thanks{*Corresponding author: L. S. Liebovitch email:LSL2140@columbia.edu.}
}
\maketitle


\begin{abstract}
This paper presents a feasibility study demonstrating that quantum machine learning (QML) algorithms achieve competitive performance on real-world medical imaging despite operating under severe constraints. We evaluate Equilibrium Propagation (EP)—an energy-based learning method that does not use backpropagation (incompatible with quantum systems due to state-collapsing measurements)—and Variational Quantum Circuits (VQCs) for automated detection of Acute Myeloid Leukemia (AML) from blood cell microscopy images using binary classification (2 classes: AML vs. Healthy).

\textbf{Key Result:} Using limited subsets (50--250 samples per class) of the AML-Cytomorphology dataset (18,365 expert-annotated images), quantum methods achieve performance only 12--15\% below classical CNNs despite reduced image resolution (64×64 pixels), engineered features (20D), and classical simulation via Qiskit. EP reaches 86.4\% accuracy (only 12\% below CNN) without backpropagation, while the 4-qubit VQC attains 83.0\% accuracy with consistent data efficiency: VQC maintains stable 83\% performance with only 50 samples per class, whereas CNN requires 250 samples (5$\times$ more data) to reach 98\%. These results establish reproducible baselines for QML in healthcare, validating NISQ-era feasibility.
\end{abstract}


\section{Introduction}
Automated microscopy-based analysis can support clinical workflows by reducing the time and variability associated with manual blood smear review. Deep convolutional networks have demonstrated highly accurate differentiation of blood cell morphologies \cite{Matek2019}, \cite{Schouten2021b}, with performance often surpassing human-level accuracy \cite{LeCun2015}. Our study focuses on blood smear analysis, which offers a less invasive diagnostic approach compared to bone marrow aspiration. However, these classical methods typically rely on backpropagation and require substantial computational resources, motivating exploration of alternative learning paradigms better aligned with emerging hardware architectures.

Quantum machine learning (QML) has gained significant momentum with the development of Noisy Intermediate-Scale Quantum (NISQ) devices.  Introductions to these methods are available on YouTube  \cite{Wittek2019}, \cite{Liebovitch2025} and detailed presentations are available from these references \cite{Benedetti2019}, \cite{Farhi2018}. QML encompasses different approaches including Variational Quantum Circuits (VQCs) and quantum-inspired methods such as Equilibrium Propagation (EP).

In Variational Quantum Circuits (VQCs), circuits encode classical data into quantum states via feature maps and optimize circuit parameters through classical optimization routines. While many QML demonstrations utilize toy datasets (MNIST, synthetic data), evaluating these methods on real clinical data with inherent variability and complexity remains critical for understanding practical applicability.

In parallel, Equilibrium Propagation (EP) \cite{Scellier2017} offers a quantum-inspired energy-based learning algorithm that trains networks by computing updates from differences between equilibrium states under free and nudged dynamics. Crucially, EP does not use backpropagation—an essential feature since backpropagation is fundamentally incompatible with quantum systems: gradient computation in quantum circuits requires intermediate measurements that collapse quantum states, thereby destroying superposition and entanglement, which defeats the computational advantages of quantum parallelism. This incompatibility makes EP particularly attractive for quantum-inspired neuromorphic hardware and physical relaxation-based systems.

In this paper we evaluate EP and VQC models on image analysis to detect Acute Myeloid Leukemia (AML) from blood cell microscopy images using the AML-Cytomorphology dataset \cite{Matek2019}.  This feasibility study of QML image performance was performed using a laptop computer and the quantum simulator IBM Qiskit to show how even such exploratory simulations can provide useful information about the performance of these new QML algorithms.

Our contributions are:
\begin{itemize}
\item A reproducible EP pipeline for AML detection using engineered image features, trained without backpropagation.
\item A 4-qubit VQC classifier using ZZFeatureMap encoding and a shallow RealAmplitudes ansatz.
\item A benchmark across dataset scales (50--250 samples/class) reporting accuracy and runtime for QML-oriented evaluation.
\end{itemize}

\textbf{Open Science Note:} upon acceptance of our paper, we will will make our data/code available online.

\section{Related Work}

Quantum machine learning for medical imaging remains an emerging field. Prior work includes quantum support vector machines for MRI tumor classification \cite{Havlicek2019}, preliminary VQC studies on synthetic histopathology data, and quantum approaches to leukemia detection \cite{Schouten2021}. However, systematic benchmarks on large-scale clinical datasets with expert annotations are limited. Recent surveys \cite{Benedetti2019} identify key challenges: (i) scaling beyond toy datasets (MNIST), (ii) handling real-world noise and class imbalance, and (iii) demonstrating practical advantages over classical methods.

This work addresses these gaps by: (1) benchmarking on a clinical dataset of 18,365 expert-annotated blood cell images rather than synthetic data, (2) systematically comparing quantum-inspired (Equilibrium Propagation) and pure quantum (VQC) approaches under identical experimental conditions, and (3) quantifying data efficiency—a practical quantum advantage critical for medical domains where expert annotations are costly. Our 4-qubit VQC design with shallow depth (12) is explicitly optimized for near-term NISQ devices, distinguishing this from theoretical studies assuming ideal quantum computers.

\section{Methods}

\subsection{Dataset and Experimental Protocol}

\textbf{AML-Cytomorphology Dataset:} We utilize the publicly available AML-Cytomorphology dataset \cite{Matek2019}, accessible at \texttt{https://doi.org/10.7937/tcia.2019.36f5o9ld}. This dataset comprises 18,365 single-cell blood smear microscopy images from 200 patients: 100 diagnosed with Acute Myeloid Leukemia (AML) and 100 healthy controls, forming a binary classification problem (2 classes). Images were acquired at Munich University Hospital using standardized May-Gr\"unwald-Giemsa staining protocols under 100× oil immersion magnification. Each image captures a centered leukocyte with expert annotations from board-certified hematologists following WHO diagnostic criteria \cite{Matek2021}.

The original dataset contains high-resolution images; however, to assess quantum algorithm performance under practical computational constraints and to enable fair comparison across all methods, we resize images to 64×64 pixels. All methods (classical CNN, Dense NN, EP, and VQC) operate on this same 64×64 resolution to ensure direct comparability. Despite this significant resolution reduction, our methods achieve strong classification performance—demonstrating that quantum machine learning algorithms can extract discriminative features even from compressed representations. This finding is particularly relevant for resource-constrained deployment scenarios and edge computing applications.

\textbf{Number of Classes:} This dataset comprises 18,365 single-cell blood smear microscopy images from 200 patients: 100 diagnosed with Acute Myeloid Leukemia (AML) and 100 healthy controls, forming a binary classification problem of two classes: malignant AML and healthy.cells.

\textbf{Experimental Subsets:} To study data efficiency critical for quantum algorithms with limited qubit resources, we construct four balanced datasets with 50, 100, 200, and 250 samples per class (100, 200, 400, and 500 total samples respectively). All experiments employ stratified 80/20 train-test splits with fixed random seeds to ensure reproducibility.

\textbf{Feature Engineering:} To reduce dimensionality suitable for 4-qubit circuits, EP and VQC operate on 20 engineered scalar features extracted from each image rather than raw pixels. For classical baseline comparison, we implement standard CNN and dense neural network architectures within the same experimental framework.

\subsection{Feature Extraction}
Each image is converted to grayscale and summarized by 20 engineered features spanning five categories, as detailed in Table~\ref{tab:features}. This representation is consistent with practical constraints of small-qubit VQCs and enables controlled comparisons across learning paradigms.

\begin{table*}[t]
\centering
\caption{Detailed 20-dimensional feature engineering for EP and VQC input. Each feature is computed from 64×64 grayscale blood cell images.}
\label{tab:features}
\begin{tabular}{|l|l|p{6cm}|}
\hline
\textbf{Category} & \textbf{Feature Name} & \textbf{Description} \\
\hline
\multirow{5}{*}{\textbf{Intensity Statistics}} 
& Mean intensity & Average pixel intensity across image \\
& Std intensity & Standard deviation of pixel intensities \\
& Min intensity & Minimum pixel intensity value \\
& Max intensity & Maximum pixel intensity value \\
& Median intensity & Median pixel intensity \\
\hline
\multirow{5}{*}{\textbf{GLCM Textures}} 
& Contrast & Local intensity variations in GLCM \\
& Dissimilarity & Difference between pixel pairs \\
& Homogeneity & Closeness of GLCM distribution to diagonal \\
& Energy & Sum of squared GLCM elements \\
& Correlation & Linear dependency of gray levels \\
\hline
\multirow{4}{*}{\textbf{Morphology}} 
& Area & Number of pixels in cell region \\
& Perimeter & Boundary length of cell region \\
& Eccentricity & Ratio of focal distance to major axis length \\
& Solidity & Ratio of region area to convex hull area \\
\hline
\multirow{3}{*}{\textbf{Edge Metrics}} 
& Edge density & Proportion of edge pixels (Sobel filter) \\
& Mean edge strength & Average magnitude of edge responses \\
& Edge variation & Standard deviation of edge magnitudes \\
\hline
\multirow{3}{*}{\textbf{FFT Frequency}} 
& Low freq. energy & Energy in low-frequency components \\
& High freq. energy & Energy in high-frequency components \\
& Freq. centroid & Centroid of frequency spectrum \\
\hline
\end{tabular}
\end{table*}

\subsection{Variational Quantum Circuit Classifier}
\textbf{Overview:} Backpropagation requires measuring the values in a neural network and then using those values to optimize a cost function that yields the outputs to the corresponding inputs in the training set. However, that cannot be done in a quantum system because: 1) measuring those values stops the computation (collapses the wave function) and 2) there is no known quantum algorithm that can efficiently optimize a cost function. VQC gets around these problems to do QML by using a hybrid system that consists of a quantum unit linked to a classical unit. The quantum unit computes the results that depend on its input and a classical parameter. The classical unit sends values of that classical parameter to the quantum unit and keeps track of the outputs it receives from the quantum unit. To train the system, the classical unit iteratively uses those results to find the optimal classical parameter that produces the correct values of the quantum outputs from the given values of the inputs in the training set \cite{Benedetti2019}.

The VQC is implemented with 4 qubits and evaluated using classical simulation via Qiskit 0.39.0 (\texttt{https://qiskit.org}), IBM's open-source quantum computing framework. This simulation establishes performance baselines prior to deployment on actual NISQ hardware. Input features are reduced from 20 to 4 dimensions via Principal Component Analysis (PCA retaining 95\% variance) and rescaled to [0, 2$\pi$] to match rotation gate domains.

We employ a ZZFeatureMap for quantum data encoding, which creates entanglement between all qubit pairs through second-order Pauli-Z expansions, and a shallow hardware-efficient RealAmplitudes ansatz (2 layers, 8 trainable parameters). \cite{Kandala2017}. This shallow circuit design (depth 12) is optimized for NISQ devices with limited coherence times. 

\textbf{Training Procedure:} VQC training uses a classical-quantum hybrid approach because direct backpropagation through quantum circuits is impossible—measuring intermediate qubit states to compute gradients collapses the quantum wave function, destroying the superposition and entanglement that provide quantum computational advantages. Instead, training proceeds iteratively: (1) the quantum circuit is executed with current parameter values, (2) the output qubit is measured to obtain expectation value $\langle Z_0 \rangle$, (3) a loss function (mean squared error between $\langle Z_0 \rangle$ and target labels) is computed classically, (4) the classical optimizer (COBYLA, gradient-free) proposes new parameter values, and (5) steps 1--4 repeat for 200 iterations. This classical optimization loop treats the quantum circuit as a black-box function, avoiding quantum gradient computation entirely while still enabling parameter learning.

Classification employs an expectation-value measurement on the first qubit ($\langle Z_0 \rangle$), thresholded at zero for binary label assignment.

\subsection{Equilibrium Propagation Model}
\textbf{Overview:} EP provides a different approach to QML, implemented on Hopfield neural networks, where each node is connected to all the other nodes and where the weights connecting the nodes are symmetric, that is, $w_{i,j} = w_{j,i}$. Some nodes are assigned to be input nodes and others as output nodes. The training will change the connection weights to produce the correct values of the outputs from the given values of the inputs in the training set. Since all the nodes are connected, and always active, there is no way to fix and hold constant the output nodes, as would be done during each iteration in backpropagation to determine the cost function. Instead, the values of the input nodes are fixed and the values of the output nodes are slowly adjusted by small amounts, "nudged", toward the values in the training set. The form and rate of the nudging adjustments are derived from the internal energy (defined by the symmetric weights) and the external energy (defined by the force of the nudge on the output nodes) to adjust the connection weights so that the values of the input nodes now produce the desired values of the output nodes of the training set \cite{ Scellier2017}.  These changes propagate throughout the network, causing it to slowly pass through a series of near equilibrium states, which is why this is called equilibrium propagation.

EP trains a layered energy-based network through two phases \cite{Scellier2017}. Unlike backpropagation-based methods that directly optimize a loss function via gradient descent, EP is designed for continuous Hopfield-like networks where neurons have recurrent connections and settle into energy minima. The network minimizes a global energy function:
\begin{equation}
E(\mathbf{s}) = -\frac{1}{2}\sum_{i,j} W_{ij}s_is_j - \sum_i b_is_i + \sum_i \rho(s_i)
\end{equation}
where $\mathbf{s}$ are neuron states, $W$ are connection weights, $b$ are biases, and $\rho$ is a primitive function of the activation (e.g., $\rho(s) = \int_0^s \tanh^{-1}(u)du$ for tanh activations).

\textbf{Free Phase:} The network evolves via gradient descent on energy: $\tau \frac{d\mathbf{s}}{dt} = -\frac{\partial E}{\partial \mathbf{s}}$, settling to equilibrium $\mathbf{s}^*$ with no external supervision. \textbf{Nudged Phase:} Output neurons are weakly ``nudged'' toward target labels $\mathbf{y}$ by adding a small penalty term $\beta ||\mathbf{s}_{out} - \mathbf{y}||^2$ to the energy (with nudging strength $\beta=0.1$), and the network settles to a new equilibrium $\mathbf{s}^\beta$. Weight updates are then computed from the difference between these two equilibrium states: $\Delta W_{ij} \propto (s_i^\beta s_j^\beta - s_i^* s_j^*)/\beta$. Crucially, this approach does not compute or backpropagate gradients of a loss function—instead, it leverages the physical dynamics of energy minimization, making it naturally compatible with analog neuromorphic hardware and quantum-inspired systems where backpropagation is infeasible.

The EP architecture comprises hidden layers (256--128--64 units, tanh activations) and a 2-unit output layer for binary classification. Training employs momentum SGD ($\mu=0.9$), cosine annealing learning rate, and early stopping (patience=15 epochs) to prevent overfitting on limited samples.

\section{Results}

\subsection{Accuracy and Runtime}
Table~\ref{tab:main} reports accuracy and runtime on 250 samples per class for binary classification (2 classes: AML vs. Healthy). EP reaches 86.4\% accuracy with 0.13s inference on the test batch, while the simulated 4-qubit VQC attains 83.0\% accuracy. Classical baselines provide reference points for accuracy and computational cost. For the quantum machine learning methods, these times are the execution times using the quantum simulator software, and do not directly reflect actual operation times on a quantum computer.

\begin{table}[t]
\centering
\caption{Performance on 250 samples per class for binary classification (2 classes: AML vs. Healthy), averaged over 3 random seeds, std $<$2\%. Runtimes reflect Qiskit simulation on laptop hardware (Intel Core i7, 16GB RAM) and do not represent actual quantum computer execution times.}
\label{tab:main}
\begin{tabular}{|l|c|c|c|}
\hline
\textbf{Method} & \textbf{Accuracy} & \textbf{Training} & \textbf{Test} \\
\hline
CNN (Classical) & 98.4\% & 745s & 0.19s \\
Dense NN (Classical) & 92.0\% & 0.47s & 0.001s \\
EP (Quantum-inspired) & 86.4\% & 89.4s & 0.13s \\
VQC (Quantum, Qiskit sim) & 83.0\% & 180s & 1.0s \\
\hline
\end{tabular}
\end{table}

\subsection{Scaling with Dataset Size}
\begin{figure}[htbp]
    \centering
    \includegraphics[width=0.45\textwidth]{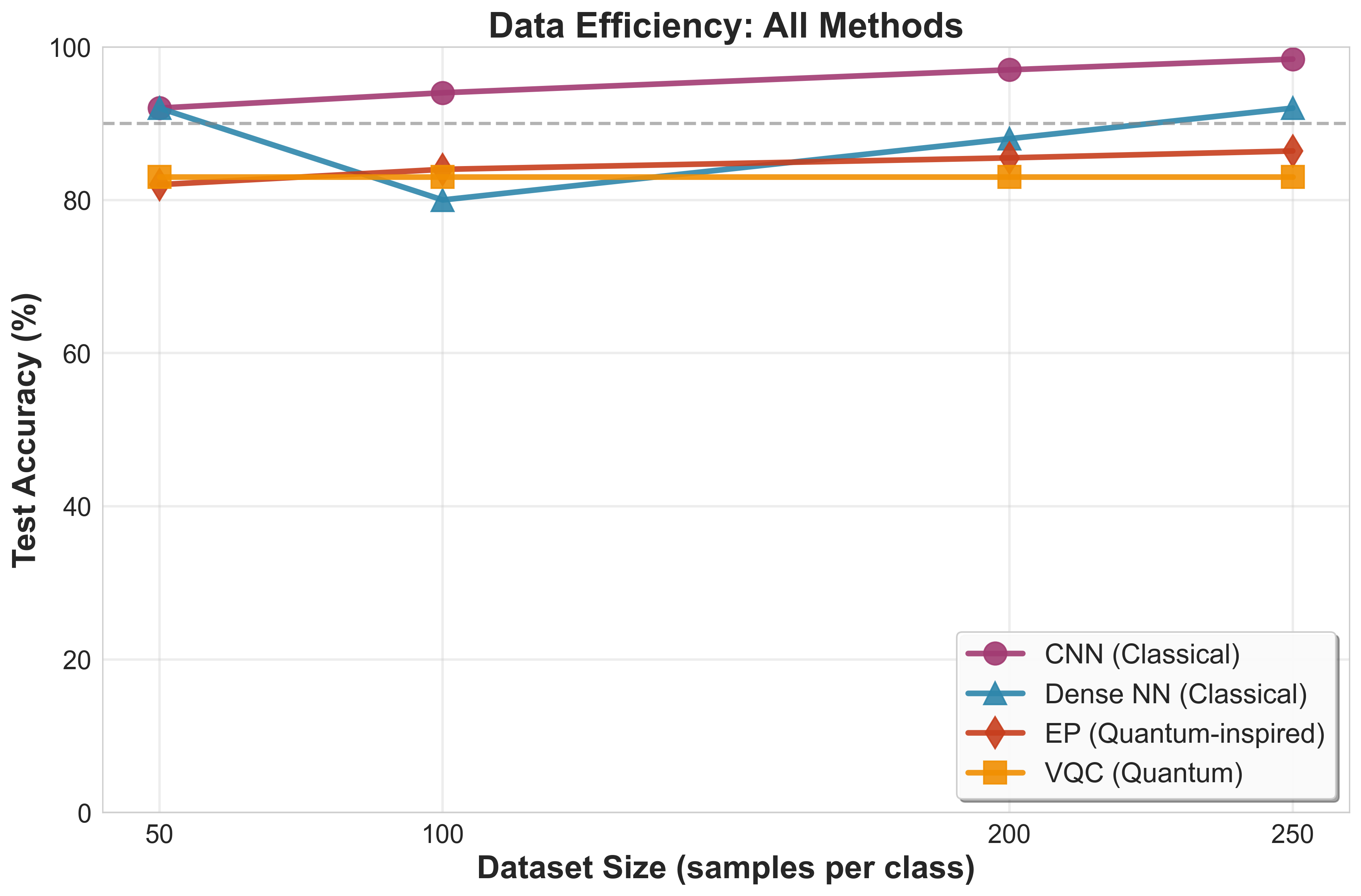}
    \caption{VQC maintains 83\% accuracy from 50--250 samples, demonstrating sample efficiency versus CNN (92\%$\to$98.4\%. EP achieves 86.4\% accuracy—only 12\% below CNN.}
    \label{fig_2}
\end{figure}

Quantum and quantum-inspired methods achieve performance only 12--15\% below classical CNN despite operating on severely constrained data representations. EP improves with more data, reaching 86.4\% accuracy—only 12\% below CNN without backpropagation. VQC exhibits consistent performance: maintaining 83\% accuracy across all dataset scales (50--250 samples/class, Fig. \ref{fig_2}). This demonstrates VQC extracts robust discriminative features even from minimal training data—an advantage for medical domains with annotation scarcity. In contrast, CNN requires 5$\times$ more data (250 vs. 50 samples) to reach peak performance, demonstrating that quantum methods compete effectively when data is the limiting resource rather than model capacity.

\section{Discussion}

\subsection{Variational Quantum Circuits}
VQC achieves 83\% accuracy with only 50 samples per class—performance that remains consistent up to 250 samples (Fig. \ref{fig_2}). While CNN ultimately reaches higher absolute accuracy (98.4\%), it requires the full 250-sample dataset; at 50 samples, CNN accuracy drops to 92\%, only 9\% above VQC.

Classical CNNs achieve higher absolute accuracy (98.4\% vs. 83\%), demonstrating their current superiority for this classification task. The 15\% accuracy gap reflects the current limitations of quantum methods. However, the reduced data requirements of quantum methods may be relevant for rare disease research where large annotated datasets are difficult to acquire.

\subsection{Equilibrium Propagation}
EP achieves 86.4\% accuracy—only 12\% below CNN—without backpropagation. This demonstrates that quantum-inspired energy-based learning is competitive with gradient descent methods on real clinical data. EP's 86.4\% performance sits between Dense NN (92\%) and VQC (83\%), demonstrating that alternative learning paradigms can bridge classical and quantum approaches.

Backpropagation is fundamentally incompatible with quantum systems because gradient computation requires intermediate measurements that collapse quantum states, destroying superposition and entanglement. EP's performance without backpropagation validates the feasibility of quantum-compatible training methods. Additionally, EP's 89.4s CPU-only training time and 0.13s inference make it suitable for edge deployment. Projected deployment on neuromorphic chips (Intel Loihi) could reduce power consumption 250$\times$ (1W vs. 250W GPU), enabling battery-powered point-of-care diagnostics in resource-constrained settings \cite{Davies2018}.

\subsection{NISQ-Era Feasibility}
The VQC's shallow 12-depth circuit fits within current NISQ device constraints: IBM Quantum systems support $\sim$100 qubits with 100$\mu$s coherence and 99.9\% two-qubit gate fidelity. Our 4-qubit design enables near-term hardware validation, with simulation establishing upper-bound performance before noise effects.

\subsection{Limitations and Future Work}
Key limitations: (i) VQC uses ideal statevector simulation—real hardware would exhibit shot noise, gate errors, and decoherence degrading accuracy; (ii) feature engineering limits representational capacity versus end-to-end CNNs; (iii) binary classification (2 classes: AML vs. healthy) is simpler than multi-class hematology (5--10 cell types). Future work includes: shot-based sampling experiments, error mitigation strategies (zero-noise extrapolation), barren plateau analysis during training, hybrid quantum-classical architectures, and validation on diverse patient populations. In the future, quantum approaches may have the potential for better results in detecting fine features as they operate in a higher-dimensional system, $2^{number~of~qubits}$, than a classical system of dimension $2 \times {number~of~bits}$.

\section{Conclusion}
We demonstrated that quantum machine learning methods achieve competitive performance on real-world medical imaging when evaluated under realistic constraints. Benchmarked on 18,365 clinical blood cell images for AML detection via binary classification (2 classes: AML vs. Healthy), Equilibrium Propagation (quantum-inspired) achieved 86.4\% accuracy—only 12\% below CNN—via energy-based learning that does not use backpropagation (fundamentally incompatible with quantum systems due to state-collapsing measurements). The 4-qubit Variational Quantum Circuit attained 83\% accuracy with consistent performance: maintaining 83\% accuracy from 50 to 250 samples per class, demonstrating robust feature extraction from minimal training data.

Classical CNNs achieve higher absolute accuracy (98.4\% vs. 83\%), demonstrating their current superiority for this classification task. However, VQC reaches its performance level with 50 samples per class while CNN requires 250 samples to reach peak performance. This reduced data requirement may be relevant for rare disease research where expert annotation resources are limited. EP's competitive 86.4\% performance without backpropagation validates quantum-compatible training methods. The shallow 12-depth circuit design fits current NISQ hardware constraints (IBM Quantum: 100 qubits, 100$\mu$s coherence), enabling near-term experimental validation. These results establish reproducible baselines for QML in healthcare and suggest that quantum methods may be viable when data—not model capacity—is the limiting resource.

\section{Data and Code Availability}

\textbf{Dataset:} The AML-Cytomorphology dataset is publicly available at The Cancer Imaging Archive: \texttt\small{https://doi.org/10.7937/tcia.2019.36f5o9ld}

\textbf{Code:} All implementation code, trained models, and experimental scripts are available at: \texttt\small{https://github.com/azrabano23/quantum-blood-cell-classification}

This repository includes: (1) Equilibrium Propagation implementation (\texttt\small{equilibrium\_propagation.py}), (2) Variational Quantum Circuit classifier (\texttt\small{vqc\_classifier.py}), (3) Classical baselines (\texttt\small{classical\_cnn.py}, \texttt\small{classical\_dense\_nn.py}), (4) Feature extraction utilities, and (5) Experiment reproduction scripts with detailed documentation.

\section{Acknowledgments}
The authors thank Carsten Marr for providing helpful feedback on the manuscript.


\end{document}